\documentclass[sigconf,natbib=true]{acmart}

\usepackage{enumitem}
\usepackage{dirtytalk}
\usepackage{subcaption}
\usepackage{textgreek}
\usepackage{multirow}
\newcommand{\padj}{p_{\mathrm{adj}}}

\newlist{questions}{enumerate}{2}
\setlist[questions,1]{label=\textbf{RQ:},ref=\textbf{RQ}}

\newtheoremstyle{boldhyp} 
    {0em}                    
    {0em}                    
    {\itshape}                    
    {\parindent}                  
    {\bfseries}           
    {.}                   
    {.3em}                
    {}                    

\theoremstyle{boldhyp}

\AtBeginDocument{%
  }

\copyrightyear{2026}
\acmYear{2026}
\setcopyright{cc}
\setcctype{by}
\acmConference[CHIIR '26]{2026 ACM SIGIR Conference on Human Information Interaction and Retrieval}{March 22--26, 2026}{Seattle, WA, USA}
\acmBooktitle{2026 ACM SIGIR Conference on Human Information Interaction and Retrieval (CHIIR '26), March 22--26, 2026, Seattle, WA, USA}
\acmPrice{}
\acmDOI{10.1145/3786304.3787872}
\acmISBN{979-8-4007-2414-5/2026/03}

\begin{document}




\title[User Perception of Digital Product Advisors Adapting to User Knowledge Levels]{``I don't know anything about laptops!'' -- User Perception of Digital Product Advisors Adapting to Their Knowledge Levels}




\author{Kevin Schott}
\email{kevin.schott@gesis.org}
\affiliation{%
 \institution{GESIS – Leibniz Institute for the Social Sciences}
 \city{Cologne}
 \country{Germany}
 }

\author{Andrea Papenmeier}
\email{a.papenmeier@utwente.nl}
\affiliation{%
 \institution{University of Twente}
 \city{Enschede}
 \country{Netherlands}
 }

\author{Daniel Hienert}
\email{daniel.hienert@gesis.org}
\affiliation{%
 \institution{GESIS – Leibniz Institute for the Social Sciences}
 \city{Cologne}
 \country{Germany}
 }

\author{Dagmar Kern}
\email{dagmar.kern@gesis.org}
\affiliation{%
 \institution{GESIS – Leibniz Institute for the Social Sciences}
 \city{Cologne}
 \country{Germany}
 }

\renewcommand{\shortauthors}{Schott et al.}

\begin{abstract}
Conversational commerce uses digital assistants to support the search process and decision-making in e-commerce. Effective communication in these interactions can be facilitated by assistants adapting their communication style to users and supporting shared understanding. An open challenge in this context is adapting the presentation of complex product information to users with varying levels of domain knowledge. To investigate strategies for such knowledge-level adaptation, we set up a chatbot-assisted laptop search scenario. In a between-subjects experiment (\(n=251\)), we examined novice and expert perceptions of product attribute recommendations presented as technical information only (T), or augmented with performance categories (TC), attribute explanations (TE), or both (TCE). For novices, approaches with explanations (TE, TCE) were perceived as more helpful and led to higher perceived learning than those without. Novices also rated the combined approach (TCE) more appropriate than the baseline (T) and TC in terms of information quantity, indicating that explanations are crucial to understand and benefit from performance categories. Critically, experts showed no significant differences across conditions, suggesting that providing supplementary information beneficial to novices did not detract from their experience. We distill these findings into four concrete design guidelines for inclusive text-based product advisors in technical domains: use TCE by default; keep a single inclusive interface; avoid standalone categories; and support user agency and personalize to the stated use case.
\end{abstract}


\acmArticleType{Research}




\maketitle


\section{Introduction}
\label{sec:introduction}
Recent advances in conversational technology have given rise to \emph{conversational commerce}, a term introduced in 2015~\cite{Messina, Tuzovic2018} and defined by Balakrishnan et al. as the ``buying activity by a customer through a digital assistant''~\cite{ConversationalCE}. In this context, users’ voice or text dialogue with a conversational agent (CA) is intended to mimic a natural exchange with a human shop assistant, gathering needs and preferences to recommend suitable products~\cite{Tsagkias2021}. By acting as an interactive, socially present decision aid, digital product advisors can foster trust in e-commerce platforms, a crucial factor for purchase intentions~\cite{Virdi2020, Gefen2003}. 

A subtle but important mechanism behind effective communication and advice is the \emph{communication accommodation theory} (CAT): interlocutors adjust their language or non-verbal cues to reduce social distance~\cite{Giles2007}. Communication accommodation not only shapes human dialogue but has also been shown to improve user experiences in human-computer interaction~\cite{Huiyang2022}. This links CAT directly to research on intelligent user interfaces and personalization, i.e., the adaptation of ``a service or a product in such a way that it fits to the preferences, cognition, requirements, or capabilities of specific persons within the confines of a specific setting''~\cite{Jbene2025}.

Prior research has demonstrated that communication accommodation strategies, such as lexical (e.g., \cite{Srivastava2023, Spillner2021, Zhao2024CUI}) and personality (e.g., \cite{Shumanov2021, Kuhail2024}) alignment, can effectively improve users' experience when interacting with CAs (see Section~\ref{subsec:personalization}). Researchers have also presented technical frameworks for persona alignment (e.g., \cite{Zhang2018, Li2016}). By contrast, knowledge-level communication adaptation--a key aspect of \textit{recipient design} (tailoring utterances to what the addressee is presumed to know and believe)~\cite{Sacks1974,Blokpoel2012}--remains comparatively underexplored, despite knowledge being a core facet of a holistic user model~\cite{Anvari2013}. Existing research has shown that LLMs can adapt the complexity of generated summaries based on user familiarity~\cite{Thakkar2024} and has presented conversational prototypes that infer user knowledge from dialogue signals to adapt responses~\cite{An2021}. However, user evaluations, especially in commercial scenarios, are scarce.

An open challenge, highlighted by prior calls for search systems in e-commerce to accommodate both low- and high-knowledge consumers (novices and experts)~\cite{Rowley2000}, is therefore to convey complex product information in an adaptive manner. Consumers with low domain knowledge tend to struggle with evaluating functional product information~\cite{Alba1987}, frequently relying on third-party opinions such as customer reviews~\cite{Ketelaar2015, Murakhovska2023}. This discrepancy not only reduces novices' ability to make informed decisions but may also impact their trust and willingness to use product advisors for complex purchases in the first place. Conversely, experts may be negatively affected by redundant information~\cite{Kalyuga2003}. While prototypes for general learning~\cite{Cai2022}, learning-oriented search~\cite{Yang2025}, and complex purchases~\cite{Murakhovska2023} have highlighted the educational potential of conversational systems, a solid understanding of how complex product information should be proactively presented to users with varying knowledge levels has not yet been established. It is, therefore, currently unclear how text-based product advisors can best adapt the \textit{type} and \textit{amount} of provided information within conversational commerce.

To address this research gap, we report on a survey-based user study integrating both quantitative and qualitative data. We examined novices' and experts' perceptions of four product attribute recommendation types in a text-based laptop advisor, varying in the type and amount of presented product information. In our experiment with 251 participants, we found that the effectiveness of supplementary information varied by information type, especially for novices. Without attribute explanations, novices reported having learned less during the interaction. They also rated the combination of \textit{performance categories} (e.g., ``mid-range configuration'') and \textit{attribute explanations} (e.g., ``The CPU [...] is a key factor in overall laptop speed and program performance.'')--condition TCE--as more appropriate in information quantity than the baseline (T) and TC. In contrast, the amount and type of supplementary information did not significantly impact how experts perceived the conversation, indicating that accommodating novices' need for additional information is not detrimental to the experience of experts.

Our findings translate into concrete design guidelines for product advisors in technical domains that can serve diverse knowledge levels. We show that novices can be effectively supported without alienating experts by supplementing technical information with attribute explanations and performance categories. Accordingly, we advocate a single inclusive interface (TCE) as the default, avoiding standalone performance categories, supporting user agency, and personalizing to the stated use case. These guidelines aim to provide a practical, empirically backed method for implementing knowledge-level recipient design.

\section{Related Work}
This section presents the theoretical frameworks that inspired our research, examines search behaviors across expertise levels, and reviews related user studies on the personalization of chatbots.

\subsection{Theoretical Frameworks}
\label{subsec:frameworks}
In human-human communication, interlocutors adapt to each other on several levels throughout the interaction. The \emph{communication accommodation theory}~\cite{Giles2007} describes how individuals adjust their communication styles to either align (convergence) with or distinguish (divergence) themselves from their conversation partners through verbal and nonverbal aspects. Previous work has established that communication accommodation is not only relevant in human-human interaction (HHI) but can also enhance the perceived interaction experience in human-computer interaction (HCI)~\cite{Huiyang2022}. Complementing the communication accommodation theory, the principle of \textit{recipient design} emphasizes that communicators actively tailor their utterances based on their hypotheses about the conversation partner's knowledge and beliefs~\cite{Sacks1974,Blokpoel2012}.

The \emph{common ground theory} states that interlocutors continuously work on establishing and maintaining a common understanding to ensure effective communication. In the HCI context, digital systems employ various grounding techniques, such as providing feedback, asking clarifying questions, and building a shared knowledge base, to mitigate errors and misunderstandings~\cite{brennan1998grounding, Tolzin2025}. Communicating understanding through active grounding also helps product advisors to be perceived as more competent~\cite{papenmeier2023ah}.

Next to these insights about communication accommodation, recipient design, and establishing common ground between interlocutors, the \emph{cognitive load theory}~\cite{Sweller2011} served as another inspiration for our research. The theory highlights the limited capacity of human working memory. It differentiates between intrinsic load (the inherent complexity of information), extraneous load (unnecessary distractions or details), and germane load (the cognitive effort to integrate new information)~\cite{Sweller2010}. In educational contexts, Hollender et al.~\cite{Hollender2010} suggested that software interfaces should adapt intrinsic load to users' domain knowledge levels and reduce extraneous load to optimize usability. This approach aligns with the \emph{expertise reversal effect}, wherein novices benefit from guided instructions, whereas extra information can burden experts, who must reconcile it with their existing knowledge structures~\cite{Kalyuga2003}. 

These theoretical frameworks can inform the implementation of \textit{adaptive selling}~\cite{Weitz1990} in e-commerce, which involves adjusting communication to the individual needs of customers to enhance satisfaction and encourage future interactions~\cite{Román2010}. Together, the concepts presented in this section directly inform the design of our study conditions (see Section~\ref{subsec:conditions}) and hypotheses (see Section~\ref{sec:hypotheses}).

\subsection{Differences in Search Behavior between Novices and Experts}
Prior work demonstrates that novice and expert users apply different search behaviors and strategies. Rowley~\cite{Rowley2000} describes that novices, characterized by lower domain knowledge, are more inclined to browse and explore information without a fixed goal, whereas experts engage in more directed searches, formulating precise queries and effectively filtering out irrelevant results. Alba and Hutchinson~\cite{Alba1987} provide additional insights into the underlying cognitive mechanisms that differentiate the two groups. They note that when experts process new product information, they draw on well-organized mental representations that enable them to retrieve relevant details, supporting informed decision-making. In contrast, novices tend to focus more on surface-level cues such as brand names and price. In terms of decision-making, experts typically interpret new data in the context of preexisting knowledge and analyze functional product attributes. Novices, in contrast, rely more on explicit product descriptions and external factors like reviews as evaluation criteria. This can render them more vulnerable to external persuasion and simple heuristic processing, such as price-quality inferences. In addition to these differences, novice consumers tend to be less confident in their purchase decisions compared to those with high domain knowledge~\cite{LetztesPaper2024}.

Collectively, these findings underscore the importance of designing (product) search systems that accommodate varying levels of user knowledge and provide adaptive support to improve novices’ search outcomes. They also highlight the importance of equipping novices with product knowledge in order to help them make informed, self-reliant decisions.

\begin{figure}
    \centering
    \includegraphics[width = 1\linewidth]{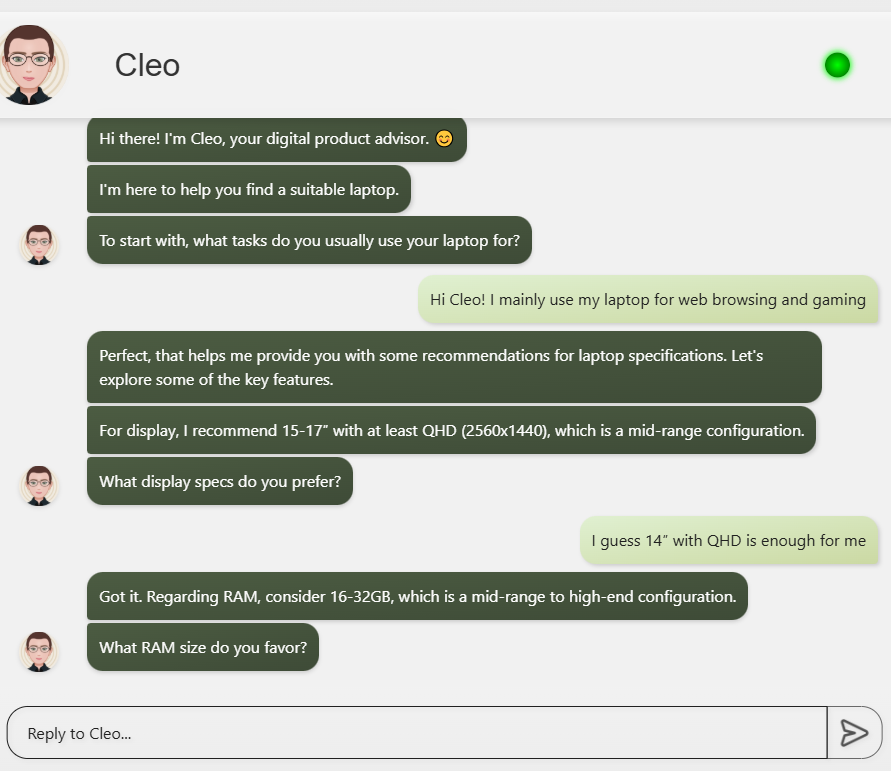}
    \caption{Interface for our product advisor ``Cleo'' (condition TC – technical information + performance categories).}
    \label{fig:Interface}
    \Description{Screenshot of a chat message exchange with our product advisor "Cleo" on a desktop device.}
\end{figure}

\subsection{Chatbot Personalization}
\label{subsec:personalization}
Personalizing interaction is a cornerstone of intelligent user interface (IUI) research. IUIs aim to represent, reason, and act on models of the user and their context~\cite{Maybury1998,Brdnik2022}. Ait Baha et al.~\cite{AitBaha2023} classify chatbot personalization into three dimensions: (i) linguistic style (e.g., vocabulary, sentence structure, tone), (ii) personal traits (e.g., personality, empathy), and (iii) persona attributes (e.g., identity, occupation, interests).

The adaptation of chatbot communication to the user, such as through lexical alignment (reusing users' wording) has proven beneficial in prior work. For instance, lexical alignment has been shown to enhance recall and understanding~\cite{Srivastava2023}, and to reduce perceived effort, frustration, mental workload, and cognitive demand~\cite{Spillner2021,Huiyang2022}. Positive trends (though non-significant) have also been observed for perceived information and service quality, enjoyment, satisfaction, and trustworthiness~\cite{Zhang2018}. Similarly, aligning chatbots' communication behavior with users' personality traits (e.g., introversion/extroversion) has been shown to improve engagement, purchasing behavior, trust, and usage intention~\cite{Shumanov2021,Kuhail2024}.

Although lexical and personality alignment are relatively well-studied, the adaptation of CAs to the user’s domain knowledge--a crucial aspect of the user model~\cite{Anvari2013}--remains comparatively underexplored. For instance, the findings of Thakkar et al.~\cite{Thakkar2024} show that LLMs can adjust the complexity of generated summaries based on user familiarity, and An et al.~\cite{An2021} aimed to implement recipient design in a conversational prototype by capturing different knowledge types identified through conversational cues. However, user evaluations of such systems are still emerging. For example, Higuchi and Inaba~\cite{Higuchi2024} demonstrated in a user study that dynamically tailoring candidate questions about news articles to users' comprehension levels can improve users' understanding compared to a baseline.

Together, these findings underscore the value of personalized chatbots. We extend this line of work by isolating knowledge-level adaptation and testing how variations in the type and amount of presented product information influence novices' and experts' perceptions in conversational commerce.

\begin{table*}[]
\small
\centering
\caption{Example attribute (RAM) recommendations for ``basic'' and ``advanced'' use cases under each condition. To isolate the effect of information presentation format, message structure and wording are held constant across use cases, and only the recommended attribute values (and corresponding performance-category labels where applicable) differ.}
\Description{For each of the four experimental conditions--T, TC, TE, and TCE--this table shows the RAM recommendation for both basic and advanced use cases.}
\begin{tabular}{p{0.175\textwidth} p{0.375\textwidth} p{0.375\textwidth}}
\toprule 
\textbf{Condition} & \textbf{``Basic'' use cases recommendation} & \textbf{``Advanced'' use cases recommendation} \\
\midrule
Technical Only (\textbf{T}) & 
    Regarding RAM, consider 8–16 GB. & 
    Regarding RAM, consider 16–32 GB. \\
\midrule
Technical + Performance Categories (\textbf{TC}) & 
    Regarding RAM, consider 8–16 GB, which is an entry‐level to mid‐range configuration. & 
    Regarding RAM, consider 16–32 GB, which is a mid‐range to high‐end configuration. \\
\midrule
Technical + Attribute Explanations (\textbf{TE}) & 
    Regarding RAM, consider 8–16 GB. RAM (working memory) helps the laptop handle multiple tasks at once and run demanding programs smoothly. & 
    Regarding RAM, consider 16–32 GB. RAM (working memory) helps the laptop handle multiple tasks at once and run demanding programs smoothly. \\
\midrule
Combined (\textbf{TCE}) & 
    Regarding RAM, consider 8–16 GB, which is an entry‐level to mid‐range configuration. RAM (working memory) helps the laptop handle multiple tasks at once and run demanding programs smoothly. & 
    Regarding RAM, consider 16–32 GB, which is a mid-range to high-end configuration. RAM (working memory) helps the laptop handle multiple tasks at once and run demanding programs smoothly. \\
\bottomrule 
\end{tabular}
\label{tab:conditions}
\end{table*}

\section{Study Design}
\label{sec:study}
Building on the identified need to understand how product advisors can effectively present complex technical information to users with varying levels of domain knowledge, our study addressed the following research question:
\begin{questions}
    \item \hypertarget{RQ}{How should text-based product advisors adapt their presentation of technical product information to meet the needs of users with different levels of domain knowledge (novices and experts)?} 
\end{questions}

To answer this research question, we conducted a controlled online user study employing a between-subjects experimental design. This approach allowed us to systematically manipulate the type and amount of supplementary information presented alongside technical product attribute recommendations, and to measure their distinct effects on novice and expert users. The study focused on the purchase of a new laptop, chosen as a representative use case of consumer electronics, an area characterized by complex technical specifications. Laptops were selected not only because of their technical complexity but also because electronic devices account for a substantial share of online purchases, representing approximately one-quarter of the global e-commerce market according to eCommerceDB (ECDB)\footnote{\url{https://ecdb.com/blog/leading-ecommerce-categories-in-top-markets/5134} last accessed: 28-09-2025}.

\subsection{Apparatus - Digital Product Advisor}
\label{subsec:apparatus}
For our experiment, we developed ``Cleo'', a text-based product advisor interface for laptop search (see Figure~\ref{fig:Interface}). To isolate the effects of the information presentation format and ensure consistent experiences across participants in each condition, Cleo was implemented as a rule-based system with a predefined interaction flow. Within this controlled structure, Cleo proactively asks users about their intended use of the laptop and their preferences for five different laptop attributes (display, RAM, storage, CPU, and GPU) in a predefined order designed to mimic the query behavior of human shop assistants~\cite{Mhm2022}. User responses regarding the intended use were categorized as either ``basic'' (e.g., text editing or web browsing) or ``advanced'' use cases (e.g., video editing, programming, or gaming). This classification determines the specific predefined technical recommendation Cleo provides for each subsequent attribute query (e.g., recommending 8-16 GB of RAM for a user with ``basic'' use cases and 16-32 GB of RAM for a user with ``advanced'' use cases). The experimental manipulation occurs in the supplementary information presented alongside these technical recommendations across conditions, as detailed in the following section.

\subsection{Information Presentation Formats (T, TC, TE, and TCE)}
\label{subsec:conditions}
We tested four information presentation formats for Cleo's laptop attribute recommendations as our experimental conditions. Each provided users with a specific type of product information or a combination of types. The following subsections describe these conditions in detail, and Table~\ref{tab:conditions} presents example messages for the attribute RAM.\footnote{The script containing the categories and explanations for all laptop attributes across all conditions is available at \url{https://osf.io/be9jt/?view_only=c7ecc97a48454e578e860bfd180ea578}.}

\subsubsection{T – Technical Information Only (Baseline)}
In condition T, the product advisor offers only numerical specifications, accompanied by standard abbreviations (e.g., QHD, RAM, SSD, GPU) for all five laptop attributes. This format serves as the baseline for all subsequent conditions, reflecting how technical specifications are typically presented in e-commerce contexts.

\subsubsection{TC – Technical + Performance Categories}
Condition TC extends condition T by adding performance categories (``entry-level,'' ``mid-range,'' ``high-end'') alongside numerical specifications. Alba and Hutchinson~\cite{Alba1987} note that categorization is part of consumers' cognitive structures, with novices often attending to basic-level categories rather than technical specifications due to limited domain knowledge. Accordingly, we add plain-language performance category labels to support novice comprehension.

\subsubsection{TE – Technical + Attribute Explanations}
In condition TE, the product advisor omits performance categories and adds a plain-language sentence explaining each attribute’s function and impact. It also replaces technical abbreviations (e.g., ``RAM'') with less specialized terms (e.g., ``working memory''). This format targets two novice needs identified by Alba and Hutchinson~\cite{Alba1987}: (i) interpreting technical specifications (via explanations and simplified terminology) and (ii) assessing relative importance (by clarifying each attribute's practical impact).

\subsubsection{TCE – Technical + Categories + Explanations}
Finally, condition TCE combines both types of supplementary information: the performance categories from TC and the attribute explanations from TE. Technical specifications and performance categories are paired with an explanatory sentence clarifying abbreviations and describing each attribute’s purpose. This format aims to maximize clarity and support informed decision-making, particularly for novice users.

\subsection{Procedure and Scenario}
We conducted the study online using SoSci Survey\footnote{\url{https://www.soscisurvey.de/en/index}}, requiring participants to access the survey via a desktop or laptop device. Participants first provided informed consent and answered closed demographic questions. They then rated their subjective domain knowledge of laptops using the item ``How would you classify your knowledge of laptops?'' on a 7-point Likert scale (1 = ``No knowledge/non-expert'' to 7 = ``High knowledge/expert'', adapted from~\cite{Brucks1985}). The use of this subjective measure is supported by findings from prior work, which demonstrated statistically significant correlations between subjective and objective knowledge across various complex domains, including technical products~\cite{Brucks1985,Raju1995} and financial investments~\cite{Goldsmith1997}. Following the demographic and knowledge assessment, participants were presented with the scenario and task description (adapted from~\cite{Papenmeier2021}):
\begin{quote}
\say{\textit{Imagine that your laptop stopped working, and you are now searching for a new one. Your task will be to answer Cleo's questions so that they can understand your needs and preferences.}}
\end{quote}

Participants were then randomly assigned and routed to one of the four interface variants (T, TC, TE, or TCE) and began interacting with Cleo, the product advisor (see Figure~\ref{fig:Interface}). Cleo sequentially asked about the participant's intended use of the laptop and preferences for the five laptop attributes (display, RAM size, storage, CPU, and GPU). Participants responded via free-text input, and Cleo returned attribute-level recommendations; no products were shown and no ranked product list was presented. After completing the interaction, participants were redirected to our questionnaire and answered several closed and one open question about their experience with Cleo. The study was approved by our institute’s ethics committee.

\subsection{Measurements}
This section outlines the dependent variables examined in the study and the instruments used to measure them.

\subsubsection{Measures for Appropriate Quantity, Relevance, Learning of Information Provided, and Trust}
These measures directly assessed key user experience aspects relevant to our research question and the theoretical framework described in Section~\ref{subsec:frameworks}. While our framing draws on cognitive load theory, we did not employ its standard measurement instruments~\cite{Babei2025} (e.g., Paas' mental-effort rating scale~\cite{Paas1992}), as they are validated for extended instructional tasks and were deemed ill-suited to our study's brief chatbot interactions. Instead, we used lightweight single-item constructs on 5-point Likert scales (1 = ``strongly disagree'' to 5 = ``strongly agree'') that are closely aligned with the aim of accommodating users' domain knowledge and are immediately evaluable post-interaction:

\begin{itemize}[leftmargin=0pt, label={}, labelsep=0pt]
    \item[] \textit{\textbf{Perceived appropriateness of information quantity}}: Whether the amount of information met user needs, using the statement ``The chatbot gave me the appropriate amount of information.'' (inspired by~\cite{Borsci2023})
    \item[] \textit{\textbf{Perceived learning}}:  Whether the interaction subjectively increased the participant’s knowledge, particularly valuable for novices, using the statement ``I learned new information about laptop aspects through the chatbot.'' (inspired by~\cite{Hiltz1994,Mello2012,Sher2009})
    \item[] \textit{\textbf{Perceived relevance}}: The extent to which the information was applicable to the user’s search, using the statement ``The information provided by the chatbot was relevant.'' (inspired by~\cite{Borsci2023})
    \item[] \textit{\textbf{Trust}}: Confidence in the chatbot, a critical factor for adoption and reliance, using the statement ``I can trust this chatbot.'' (inspired by~\cite{Virdi2020,Gefen2003})
\end{itemize}

\subsubsection{Measures for Perceived Helpfulness}
We assessed the perceived helpfulness of the supplementary information to determine the specific utility of each information component (performance categories, attribute explanations, or both in combination), relating to the practical value of the provided adaptations. In conditions TC, TE, and TCE, participants were shown screenshots of two example attribute recommendations for their assigned condition. They rated the statement ``How helpful did you find the [categorization/explanation/supplementary information] for understanding the recommendation?'' on a 5-point Likert scale ranging from 1 = ``Not at all helpful'' to 5 = ``Very helpful.''

\subsubsection{Attribute Recommendation Accuracy}
To evaluate perceptions of the core technical recommendations, participants rated their accuracy on a 5-point Likert scale (1 = ``Not at all accurate'' to 5 = ``Totally accurate'') with the residual answer option ``I don't know.'' This measure ensured that the perceived plausibility of Cleo’s recommendations did not confound participants’ evaluations of the supplementary information.

\subsubsection{Qualitative Feedback}
Finally, participants responded to an open-text question: ``What additional information would have helped you to answer Cleo's questions about your preferences?'' This qualitative feedback provided insight into how novices and experts perceived the supplementary information presented under the different conditions.

\subsection{Hypotheses}
\label{sec:hypotheses}
This section presents our hypotheses regarding the perceptions of novices and experts, both across and within experimental conditions. Our hypotheses are broadly guided by the assumption that novices, who often struggle with unfamiliar technical product information, will benefit from supplementary information~\cite{Alba1987}. Conversely, experts may find such information redundant or less impactful, aligning with concepts like the \emph{expertise reversal effect}~\cite{Kalyuga2003}. We therefore expect divergent responses to variations in the type and amount of information provided, depending on user expertise.

For \textit{perceived appropriateness}, \textit{learning}, \textit{relevance}, and \textit{helpfulness}, we generally expect novices to prefer more comprehensive information, while experts' perceptions will be less influenced by supplementary details, or they may prefer less:

\begin{itemize}[topsep=5pt]
    \item[\textbf{H1a:}] Novices deem the combination of performance categories and attribute explanations (TCE) as most appropriate, while each type of supplementary information alone (TC, TE) is rated as more appropriate than no supplementary information (T).
    \item[\textbf{H1b:}] Experts perceive no supplementary information (T) as most appropriate.
    \item[\textbf{H1c:}] Novices rate conditions with supplementary information (TC, TE, TCE) higher in appropriateness than experts.
    \item[\textbf{H1d:}] Experts perceive the no-supplementary-information condition (T) as more appropriate than novices do.
    
    \addvspace{0.7\baselineskip}

    \item[\textbf{H2a:}] Novices experience the highest perceived learning with TCE, while TC and TE lead to more perceived learning than T.
    \item[\textbf{H2b:}] Experts show no significant differences in perceived learning between conditions.
    \item[\textbf{H2c:}] Novices report higher perceived learning in conditions TC, TE, and TCE compared to experts.

    \addvspace{0.7\baselineskip}

    \item[\textbf{H3a:}] Novices perceive TCE as most relevant, while TC and TE are deemed more relevant than T.
    \item[\textbf{H3b:}] Experts perceive no supplementary information (T) as most relevant.
    \item[\textbf{H3c:}] Novices perceive conditions TC, TE, and TCE as more relevant than experts do.
    \item[\textbf{H3d:}] Experts find condition (T) more relevant than novices.

    \addvspace{0.7\baselineskip}

    \item[\textbf{H5a:}] Novices perceive the performance categories combined with attribute explanations (TCE) as most helpful.
    \item[\textbf{H5b:}] Experts show no significant differences in perceived helpfulness across conditions TC, TE, and TCE.
    \item[\textbf{H5c:}] Novices rate conditions TC, TE, and TCE as more helpful than experts.
\end{itemize}

Finally, building on Kuhail et al.'s finding that chatbot personality alignment can enhance \textit{trust}~\cite{Kuhail2024}, we hypothesize that alignment with a user’s domain knowledge will have a similar effect:

\begin{itemize} [topsep=5pt]
    \item[\textbf{H4a:}] Novices show most trust with TCE, while TC and TE enhance novices' trust compared to T.
    \item[\textbf{H4b:}] Experts show increased trust with no supplementary information (T).
    \item[\textbf{H4c:}] Novices exhibit higher trust in conditions TC, TE, and TCE than experts.
    \item[\textbf{H4d:}] Experts show higher trust than novices in condition T.
\end{itemize}

\section{Participants}
\label{subsec:participants}
We recruited participants on Prolific\footnote{\url{https://www.prolific.com/}} with eligibility criteria requiring residence in the UK or USA, English as their primary language, and age $\geq 18$ years. To ensure data quality, we required a Prolific approval rate of at least 95\% and a minimum of 20 previous submissions, and we included an attention-check question~\cite{Oppenheimer2009}. One participant was excluded for being under 18, yielding a final sample of 251 participants (124 male, 126 female, one preferred not to disclose; age 18--77, $M=41.2,\, SD=13.6$).

Following prior work that classifies consumers by subjective domain knowledge (e.g., \cite{Nam2012, Vigar2015, Lee2016}), we defined novices as ratings 1--4 and experts as ratings 5--7 on the 7-point subjective domain-knowledge scale before data collection. The threshold was specified before data collection based on an archival dataset collected with the same item in a comparable population (participants recruited via Prolific, located in the UK, and speaking English as their primary language)\footnote{The anonymized dataset is available at \url{https://osf.io/be9jt/?view_only=c7ecc97a48454e578e860bfd180ea578}.}. In that dataset (\(N=135\)), counts by scale point (1--7) were 3, 17, 14, 32, 37, 22, and 10, yielding grouped totals of 1--4$\,=\,$66 (48.9\%) and 5--7$\,=\,$69 (51.1\%), which informed our recruitment quotas. Using stratified random assignment, participants from each stratum (novice/expert) were assigned with equal probability to one of four conditions (T, TC, TE, TCE). Recruitment continued until the stopping rule was met: at least 30 novices and 30 experts in each condition. The resulting groups showed distinct means on the 1--7 scale: novices ($M=3.0,\, SD=1.0$) and experts ($M=5.6,\, SD=0.8$). Final per-condition counts were: novices (T = 30, TC = 30, TE = 30, TCE = 35) and experts (T = 31, TC = 30, TE = 35, TCE = 30).

Participants took part in the study for an average of 6:40 minutes ($SD =$ 4:13 minutes) and received 1.20~GBP in compensation, equivalent to an hourly rate of approximately 10~GBP.

\section{Results}
This section presents our findings for the user study. First, we describe the results of our statistical analyses and evaluate our hypotheses. We then present the qualitative insights we gathered from the participants' responses to our open question.

\subsection{Statistical Analysis}
\begin{figure*}
  \centering
  \makebox[\textwidth][c]{%
    \subfloat[Perceived appropriateness of information quantity]{%
      \includegraphics[width=0.34\textwidth]{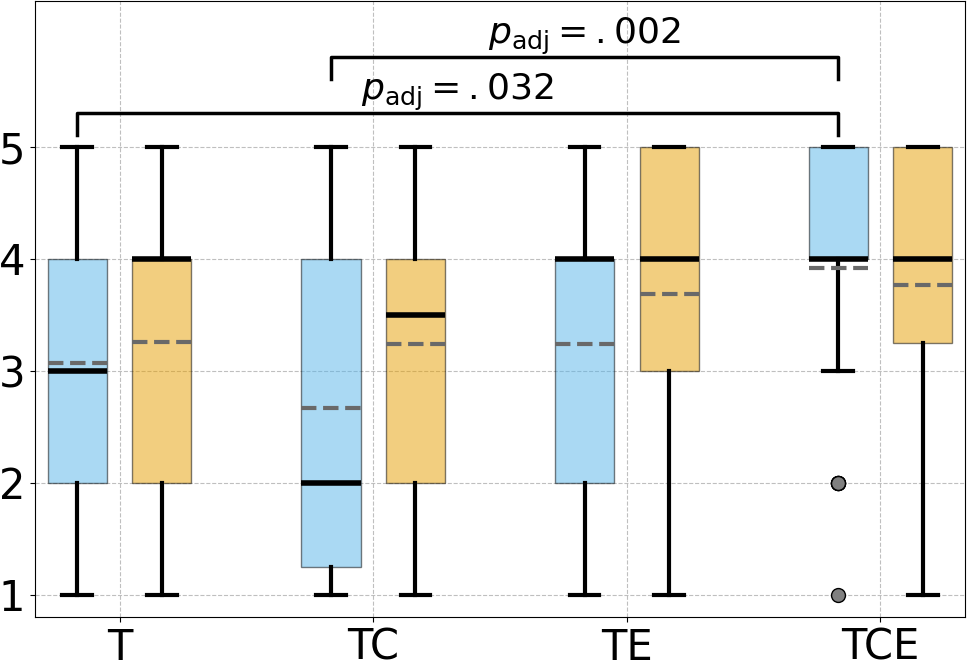}%
      \label{PT1}%
      \Description{Box plots visualizing the distribution of novices' and experts' responses for the variable perceived appropriateness of information quantity for each of the four conditions. Mean and standard deviation values are provided in Table 2, row 1. Brackets above the box plots indicate that novices rated TCE significantly higher than both T and TC. No additional significant differences between conditions are indicated.}%
    }%
    \hspace{3mm}%
    \subfloat[Perceived learning]{%
      \includegraphics[width=0.34\textwidth]{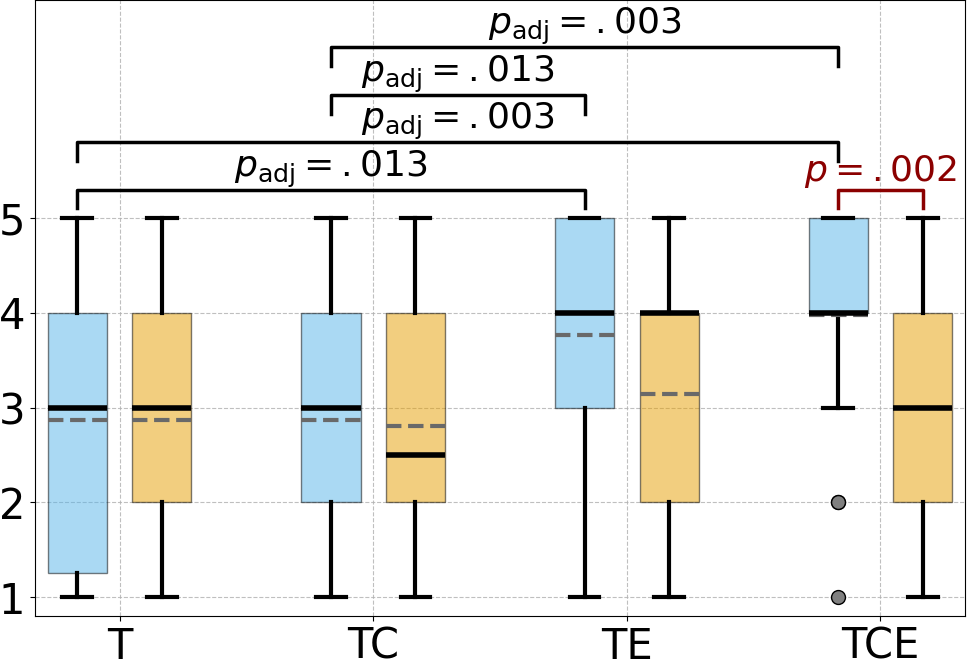}%
      \label{PT2}%
      \Description{Box plots visualizing the distribution of novices' and experts' responses for the variable perceived learning of information provided for each of the four conditions. Mean and standard deviation values are provided in Table 2, row 2. Brackets above the box plots indicate that novices rated both TE and TCE significantly higher than T and TC. An additional bracket indicates that novices rated the condition TCE significantly higher than experts. For experts, no significant differences between conditions are indicated.}%
    }%
  }
  \\[2mm]
  \makebox[\textwidth][c]{%
    \subfloat[Perceived relevance]{%
      \includegraphics[width=0.34\textwidth]{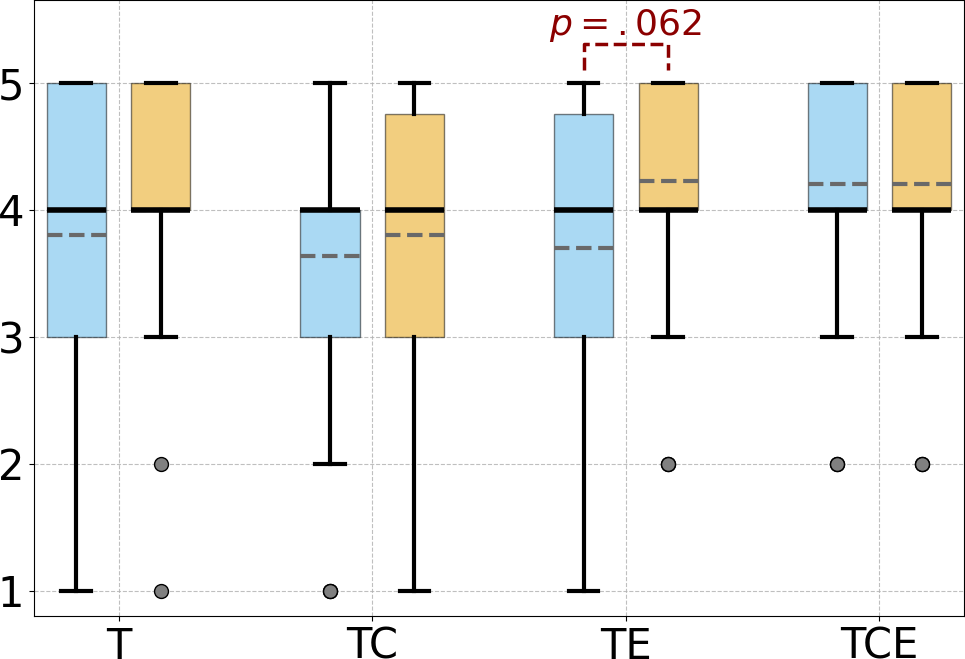}%
      \label{PT3}%
      \Description{Box plots visualizing the distribution of novices' and experts' responses for the variable perceived relevance of information provided for each of the four conditions. Mean and standard deviation values are provided in Table 2, row 3. No significant differences are indicated.}%
    }%
    \hspace{3mm}%
    \subfloat[Trust]{%
      \includegraphics[width=0.34\textwidth]{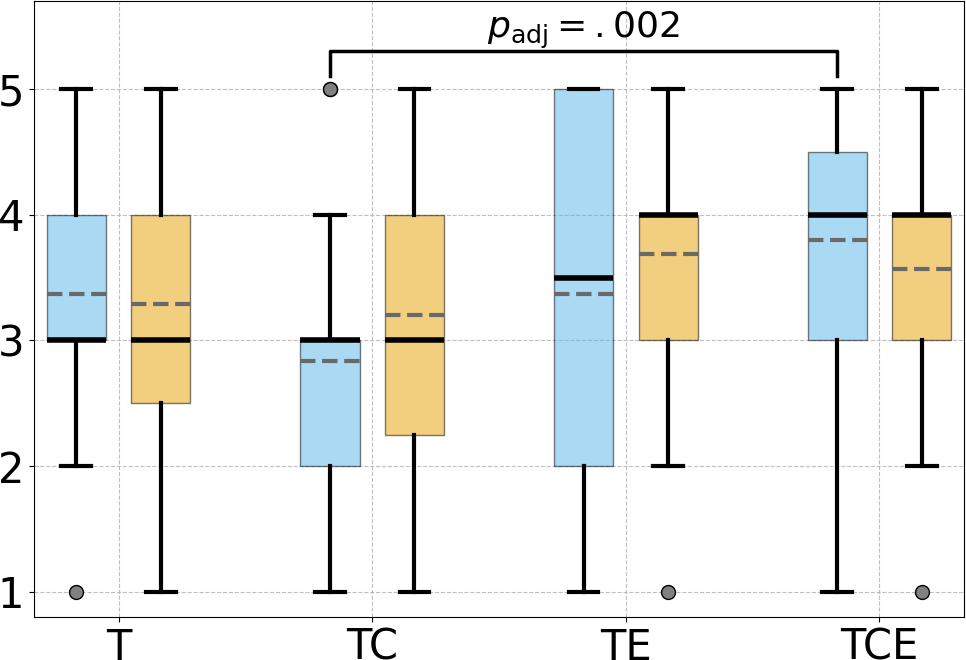}%
      \label{PT4}%
      \Description{Box plots visualizing the distribution of novices' and experts' responses for the variable trust for each of the four conditions. Mean and standard deviation values are provided in Table 2, row 4. A bracket above the boxplots indicates that novices rated TCE significantly higher than TC. No additional significant differences between conditions are indicated.}%
    }%
    \hspace{3mm}%
    \subfloat[Perceived helpfulness]{%
      \includegraphics[width=0.29\textwidth]{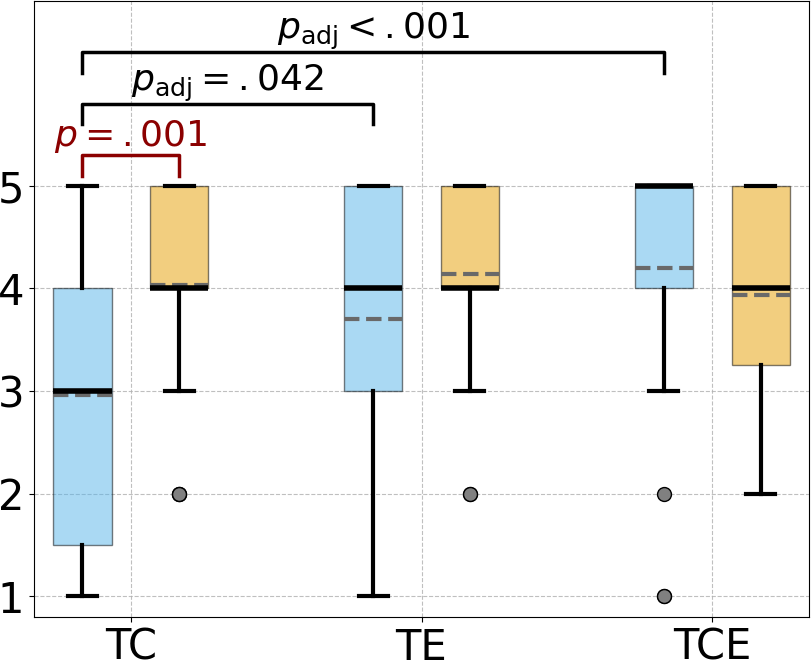}%
      \label{PT5}%
      \Description{Box plots visualizing the distribution of novices' and experts' responses for the variable perceived helpfulness of information provided for each of the four conditions. Mean and standard deviation values are provided in Table 2, row 5. Brackets above the box plots indicate that novices rated both TCE and TE significantly higher than TC. An additional bracket indicates that experts rated the condition TC significantly higher than novices. For experts, no significant differences between conditions are indicated.}%
    }%
  }
  \caption{Box plots comparing novices' (blue) and experts' (orange) ratings across our dependent variables. Black horizontal lines represent medians, and dashed grey lines indicate means. Black and red brackets indicate significant differences determined by Dunn’s post-hoc (\(\padj<.05\)) and Mann–Whitney U tests (\(p<.05\)), respectively.}
  \label{fig:boxplots}
\end{figure*}

Because our data was not normally distributed, we used Kruskal-Wallis tests with post-hoc Dunn's tests to test for significant differences between our four conditions among novices and among experts. Because this analysis involves multiple pairwise post-hoc comparisons per dependent variable, we controlled for multiple testing using the Benjamini-Hochberg procedure with a false discovery rate of \(q=.05\). Unlike Bonferroni-style family-wise error control, this approach limits the expected proportion of false positives among the set of significant findings while retaining more statistical power. We applied BH correction within each family of post-hoc comparisons per dependent variable and expertise group. Accordingly, we report BH-adjusted post-hoc \(p\)-values, denoted \(\padj\). For novice-expert comparisons within a condition, we used Mann-Whitney U tests. To quantify effect sizes for non-parametric tests, we report Cliff's Delta (\(\delta\))~\cite{Wan2021}. Means and standard deviations for each dependent variable are presented in Table~\ref{tab:NovicesExperts}. Meanwhile, Figure~\ref{fig:boxplots} visualizes our results and indicates significant differences.

\begin{table*}[h]
\centering
\caption{Means (standard deviations) for each dependent variable (scales ranging from 1 = ``strongly disagree'' to 5 = ``strongly agree'') by condition, separated for novices and experts.}
\Description{This table presents mean scores (with standard deviations in parentheses) for our five dependent variables--appropriateness, learning, relevance, trust, and helpfulness--across our four experimental conditions (T, TC, TE, and TCE). For each condition, ratings are provided separately for novices and experts. Perceived helpfulness only contains data for the conditions TC, TE, and TCE.}
\label{tab:NovicesExperts}
\small
\begin{tabular}{|p{2.1cm}|c|c||c|c||c|c||c|c|}
\hline
                         & \multicolumn{2}{c||}{\textbf{T}}       & \multicolumn{2}{c||}{\textbf{TC}}      & \multicolumn{2}{c||}{\textbf{TE}}      & \multicolumn{2}{c|}{\textbf{TCE}}      \\
\cline{2-9}
                         & Novices & Experts         & Novices & Experts         & Novices & Experts         & Novices & Experts         \\
\hline
\textbf{Appropriateness}          & 3.10 (1.35) & 3.26 (1.15)  & 2.67 (1.45) & 3.25 (1.43)  & 3.31 (1.35) & 3.69 (1.13)  & 3.91 (1.21) & 3.96 (1.14)  \\
\textbf{Learning}                 & 2.86 (1.41) & 2.87 (1.34)  & 2.87 (1.36) & 2.86 (1.21)  & 3.73 (1.28) & 3.14 (1.44)  & 3.97 (0.95) & 3.20 (1.22)  \\
\textbf{Relevance}                & 3.79 (1.15)         & 4.00 (0.97)  & 3.63 (1.13)         & 3.86 (1.04)  & 3.88 (1.03)         & 4.23 (0.88)  & 4.20 (0.90)         & 4.36 (0.86)  \\
\textbf{Trust}                    & 3.38 (0.90) & 3.29 (1.07)  & 2.83 (1.05) & 3.18 (1.25)  & 3.46 (1.39) & 3.69 (0.96)  & 3.80 (1.05) & 3.76 (1.01)  \\
\textbf{Helpfulness}             & --                  & --           & 2.97 (1.35) & 4.04 (1.00)  & 3.88 (1.21) & 4.14 (0.91)  & 4.20 (1.11) & 3.92 (1.15)  \\
\hline
\end{tabular}
\end{table*}

\subsubsection{Perceived Appropriateness of Information Quantity}
A Kruskal-Wallis test revealed a significant difference among the four conditions for novices (\(H=13.736,\, p=.003\)). Dunn’s post hoc tests indicated that novices perceived TCE as significantly more appropriate than TC (\(\padj=.002,\, \delta=.478\), large effect), and more appropriate than T (\(\padj=.032,\, \delta=.372\), medium effect). The differences between TC and T, TE and T, TE and TC, and TCE and TE were not significant (all \(\padj>.05\)), thereby \textbf{only partly supporting H1a}. For experts, no significant differences emerged among the four conditions (\(H=4.926,\, p=.177\)), thus \textbf{rejecting H1b}. Mann–Whitney U tests comparing novices and experts within each condition revealed no significant differences (all \(p>.05\)), thus also \textbf{rejecting H1c} and \textbf{H1d}.

\subsubsection{Perceived Learning}
Among novices, the Kruskal-Wallis test showed a significant effect of condition on perceived learning (\(H=18.525,\, p<.001\)). Dunn’s post hoc tests revealed that both TE and TCE were rated significantly higher than T (TE vs. T: \(\padj=.013,\, \delta=.371\); TCE vs. T: \(\padj=.003,\, \delta=.459\)) and higher than TC (TE vs. TC: \(\padj=.013,\, \delta=.381\); TCE vs. TC: \(\padj=.003,\, \delta=.471\)); all reflecting medium effect sizes. TE and TCE did not differ (\(\padj=.693\)), thereby \textbf{partly supporting H2a}. For experts, no significant differences were observed among the four conditions (\(H=1.335,\, p=.721\)), \textbf{supporting for H2b}. When comparing novices and experts within each condition using Mann–Whitney U tests, novices rated TCE significantly higher (\(U=752.5,\, p=.002,\, \delta=.433\), medium effect), while no other comparisons reached significance (all \(p>.05\)), thereby \textbf{partly supporting H2c}.

\subsubsection{Perceived Relevance}
Kruskal-Wallis tests for both novices (\(H=5.916,\, p=.116\)) and experts (\(H=4.769,\, p=.190\)) revealed no significant differences among the four conditions. Mann–Whitney U tests comparing novices and experts within each condition also showed no significant differences, although TE exhibited a potential trend toward significance (\(U=392.5,\, p=.062,\, \delta=.252\), small effect). No other comparisons approached significance, \textbf{rejecting H3a}, \textbf{H3b}, \textbf{H3c}, and \textbf{H3d}.

\subsubsection{Trust}
For novices, the Kruskal-Wallis test indicated a significant difference across conditions (\(H=12.713,\, p=.005\)). Dunn’s post hoc tests showed that TCE was rated significantly higher than TC (\(\padj=.002,\, \delta=.500\), large effect), while no other pairwise comparisons were significant, thereby \textbf{partly supporting H4a}. In contrast, experts showed no significant differences among the four conditions (\(H=4.308,\, p=.230\)), \textbf{rejecting H4b}. Mann–Whitney U tests revealed no significant novice–expert differences within any condition (all \(p>.05\)), \textbf{rejecting H4c} and \textbf{H4d}.

\subsubsection{Perceived Helpfulness (only for TC, TE, and TCE)}
A Kruskal-Wallis test among novices (\(H=15.989,\, p<.001\)) showed a significant difference in perceived helpfulness across TC, TE, and TCE. Dunn’s post hoc tests indicated that novices found both TE and TCE significantly more helpful than TC (TE vs. TC: \(\padj=.042,\, \delta=.319\), small effect; TCE vs. TC: \(\padj<.001,\, \delta=.553\), large effect), while TE and TCE did not differ significantly (\(\padj=.086,\, \delta=.240\), small effect), thereby \textbf{partly supporting for H5a}. Experts showed no significant differences among TC, TE, and TCE (\(H=.484,\, p=.785\)), \textbf{supporting H5b}. Mann–Whitney U tests further revealed that experts rated TC as significantly more helpful than novices (\(U=240.5,\, p=.001,\, \delta=.465\), medium effect), while no other novice–expert comparisons were significant (all \(p>.05\)), thus \textbf{rejecting H5c}.

\subsubsection{Perceived Accuracy of Attribute Recommendations}
Both novices ($M=3.83,\, SD=0.86,\, n=72$, 53 ``I don't know'') and experts ($M=3.94,\, SD=0.85,\, n=113$, 13 ``I don't know'') perceived the attribute recommendations provided by the product advisor as fairly accurate.

\subsection{Qualitative Feedback}
To analyze participants' responses to our open-ended question, ``What additional information would have helped you to answer Cleo's questions about your preferences?'', we performed a two-stage thematic analysis~\cite{Braun2006}. First, an independent researcher uninvolved in the study design and data collection conducted an initial inductive coding (i.e., generating initial codes) on all participants' responses in close discussion with the first author. This stage collaboratively developed the coding scheme and produced a final codebook.\footnote{The full codebook is available at \url{https://osf.io/be9jt/?view_only=c7ecc97a48454e578e860bfd180ea578}.} Subsequently, the independent researcher and the first author independently recoded all responses using the finalized scheme. Responses could receive multiple codes when they encompassed several distinct themes. The coders then reconciled disagreements through discussion until consensus, following negotiated-agreement best practices~\cite{McDonald2019}.

The qualitative feedback revealed clear divergences between novices (\(n=125\)) and experts (\(n=126\)). The most significant difference was in \textit{Attribute Confusion}, where participants struggled to understand technical terms, what each attribute does or is relevant for, and asked for further explanations. For example, P17 (novice in condition TC), expressed, \emph{``If [Cleo] could have given me more information about the different computer terms that [they were] using.''}. This was the most mentioned theme by novices (49\% overall), most pronounced in conditions T (60\%) and TC (60\%), with lower frequencies in TE (43\%) and TCE (34\%). For experts, this was a minor issue (15\% overall), with the highest rate appearing in condition TC (23\%), followed by T (16\%), TE (11\%), and TCE (10\%). Conversely, experts’ primary request was for \textit{Missing Specs \& Attributes}--indicating a desire to specify or receive information on additional attributes like price, brand, or weight--(36\% overall), with frequencies highest in conditions TE and TCE (both 43\%), compared to T (32\%) and TC (23\%). This theme was far less frequent among novices (12\% overall), where it was highest in condition TE (17\%), followed by TCE (14\%) and TC (13\%), and was rarest in T (3\%). \textit{P235} (expert in condition TC), for example, stated, \emph{``Maybe give me some brand names of those GPUs or CPUs''}. Other differences were smaller: only experts mentioned needing \textit{Improved Recommendations} (suggestions that the product advisor’s attribute recommendations should be refined, 4\%), while only novices called the chatbot a \textit{Non-preferred Medium} (and preferring other mediums such as going to a physical store or reading online reviews, 2\%). Finally, more than twice as many experts as novices reported that they received enough information and would not have needed additional information compared to novices (13\% vs. 6\%). For example, \textit{P71} (expert in condition TC) noted: \emph{``Nothing needed as I am very comfortable in laptop technology''}.

Despite these differences, a key commonality was the desire for a \textit{Provision of Options}--where novices noted they would have liked to be provided more than one attribute value recommendation to choose from or a comparison of different value options. \textit{P77} (novice in condition TC) articulated, \emph{``Giving more options for a given category (RAM, GPU, Storage, etc.) and not just what Cleo thinks is the best option''}. Overall, this theme was mentioned by 14\% of novices and 17\% of experts. For novices, this desire was highest in conditions TC (20\%) and TE (17\%), and lower in TCE (11\%) and T (10\%). For experts, rates were highest in TE and TCE (20\% each) and lower in T and TC (13\% each). Similarly, both groups expressed a desire for more \textit{Personalization} (i.e., tailoring the advisor's questions and explanations to a user's specific use cases, 13\% for experts vs. 8\% for novices) and better \textit{Interaction Quality} (such as wanting to ask questions, have a more natural conversation, or have the system confirm its understanding, 14\% for novices, 11\% for experts).

\section{Discussion \& Implications}
Our findings reveal a consistent picture of the novice experience, where the type of supplementary information provided is critical. Novices rated conditions with attribute explanations (TE and TCE) as significantly more helpful than performance categories alone (TC), and they found the combination of both (TCE) significantly more appropriate than the baseline (T) and TC. Qualitative data provides insight into this preference: the theme of \textit{Attribute Confusion} was the major concern for novices, voiced by 60\% in the explanation-free baseline (T) and TC conditions. This confusion was mentioned only around half as frequently in the TCE condition, suggesting that novices must first understand an attribute's function and impact before they can benefit from the more abstract performance categories–an interpretation also supported by their significantly higher reported learning in TCE versus TC. Critically, our findings suggest that designing for novices must not come at the expense of the expert experience: experts showed no significant perceptual differences across conditions. Their qualitative feedback indicates that they had different priorities, with their primary request being for \textit{Missing Specs \& Attributes} like brand and price, which suggests the supplementary information was not as relevant to their evaluation.

These results can be viewed through several theoretical lenses. From a \textit{communication accommodation theory}~\cite{Giles2007} and \textit{recipient design}~\cite{Sacks1974,Blokpoel2012} perspective, our findings can be interpreted as an instance of effective knowledge-level accommodation. This form of persona-based personalization for technical products appears to create more suitable interactions for novices, in line with prior work showing positive effects of communication accommodation in chatbots~\cite{Srivastava2023, Spillner2021, Shumanov2021, Kuhail2024}. Our findings may also relate to \textit{common ground theory}~\cite{brennan1998grounding}: providing explanations (TE and TCE) seems to have facilitated the establishment of a knowledge base in novices, as indicated by their significantly higher perceived learning compared to conditions T and TC. While Cleo's rule-based nature limited dynamic grounding, this initial knowledge-building is a key step toward a shared user-system understanding~\cite{Tolzin2025}. Additionally, while we did not measure cognitive load directly, our findings indicate a nuanced view of the \textit{expertise reversal effect}~\cite{Kuhail2024} and \textit{cognitive load theory}~\cite{Sweller2011} as inspirations for our hypotheses. While novices seem to have benefited from the supplementary information in TCE as expected, it seems this information did not concern experts, conflicting with our assumptions. The concise, sequential information may have allowed experts to efficiently filter or skim content not immediately relevant to them, served as quick confirmations rather than distractions, or still offered some utility to them, preventing the adverse extraneous cognitive load the expertise reversal effect would predict.

Beyond our primary hypotheses, the qualitative analysis revealed shared desires that our quantitative metrics did not capture. Both groups requested more personalized recommendations and the ability to compare different attribute values rather than receiving a single recommendation. Notably, both novices and experts mentioned personalization most frequently in condition TE (13\% novices vs. 20\% experts), suggesting that while participants rated the chatbot's supplementary information as broadly relevant across all conditions (with no significant differences found), attribute explanations provided objective information that users would have liked to be catered to their stated use cases. Meanwhile, participants' desire for greater control and agency in their decision-making process reflects established HCI principles (e.g., \cite{Nielsen1994,Shneiderman2017}).

Based on these insights, we propose the following design guidelines for text-based product advisors in technical domains:

\begin{enumerate}[leftmargin=*, itemsep=2pt, topsep=2pt, label=\textsc{\arabic*}.]
  \item \textbf{Default to TCE as information presentation format:} technical specs + performance categories + attribute explanations.
  \item \textbf{Use one inclusive interface:} don’t hide supplementary information for experts.
  \item \textbf{Do not present performance categories in isolation:} pair performance categories with an attribute explanation.
  \item \textbf{Support user agency \& personalization:} offer options, let users add missing specs, and tailor explanations to the stated use case.
\end{enumerate}

\section{Limitations}
\label{sec:limitations}
Several limitations should be considered when interpreting our findings. First, our study focused on the domain of laptop search. Different accommodation and explanation strategies may be more suitable in other domains, suggesting a need for further domain-specific investigations of knowledge adaptation approaches.

Second, we operationalized expertise using an a priori threshold and used stratified random assignment. Although novices and experts differed substantially in mean knowledge scores and condition-specific responses--suggesting that the categorization was fit for purpose--the approach relies on self-reports, which can be biased~\cite{Karpen2018}, and binary thresholds on continuous measures entail information loss and reduced statistical efficiency~\cite{DeCoster2011}. We therefore treat ``novice'' and ``expert'' as coarse groupings for design and exposition. Future work should supplement self-reports with objective, standardized knowledge tests (e.g., laptop-component quizzes) and analyze domain knowledge as a continuous moderator to obtain more granular inference. The limitation of self-reporting also applies to our measurement of learning.

Finally, the experimental design employed a rule-based advisor and scripted prompts to isolate presentation effects. This control precluded free-form, mixed-initiative dialogue and limited opportunities for dynamic grounding~\cite{Clark1991}, thereby reducing ecological validity. Future studies should replicate the investigation in more open conversational settings.

\section{Future Work}
Our research opens several promising directions. First, future work could explore alternative communication accommodation strategies, such as varying formality or message length, and examine how our findings translate to voice-based and more interactive conversational agents. The latter would also enable analysis of the dynamic establishment of common ground, e.g., via conversation analysis (e.g., \cite{Convertino2008}) or by measuring grounding costs (e.g., \cite{Homaeian2021}). Furthermore, developing and evaluating methods for real-time user assessment could enable dynamic adaptation of explanatory depth within a single inclusive interface--scaling detail in response to explicit requests, confusion detected through behavioral or physiological signals, or knowledge inferred from conversational cues (building on~\cite{An2021}) while retaining the minimal explanation that accompanies any category. Our presentation-format findings could also inform prompt engineering or instruction tuning for LLM-based product advisors, where domain-knowledge signals guide adaptive explanation depth.

Second, several design alternatives merit investigation. One promising approach is progressive elaboration: present any performance category with a concise one-sentence explanation, and make additional details accessible via tooltips or expandable message bubbles. This may be especially valuable when users’ expertise varies across attributes. Another avenue is to integrate multimedia (e.g., images, videos) to support supplementary information and facilitate learning for users.

Third, future studies could compare user preferences for system guidance versus user autonomy. Analyzing how novices and experts differ in their preference for a question-driven guided interaction versus a direct-query model, as well as their query behavior, could inform the design of more flexible, mixed-initiative systems.

\section{Conclusion}
Novices in e-commerce often lack sufficient domain knowledge to evaluate complex products by attributes, relying instead on third-party opinions. Conversational commerce offers the opportunity to adapt to users' knowledge levels, provide educational value, and thus support informed decision-making. Our research, therefore, examined how different combinations of technical information, performance categories, and attribute explanations influence user perception in a laptop search scenario.

Our findings demonstrate that approaches including attribute explanations (TE and TCE) were perceived as significantly more helpful and enhanced perceived learning for novices, who perceived the combined approach (TCE) specifically as more appropriate than the baseline (T) and TC in terms of information quantity. This indicates the need for the right kinds of supplementary information to help bridge novices' knowledge gaps. Crucially, we did not detect a significant negative impact on expert users, who showed no significant preference across conditions.

We distill these findings into four design guidelines for text-based product advisors in technical domains: use TCE by default; stick to one inclusive interface; do not present performance categories without an explanation; and preserve user agency (by offering options and letting users add missing specs) and personalize to the stated use case. Together, these guidelines aim to operationalize knowledge-level recipient design for novices without alienating experts. Future research should continue exploring this strategy across different product domains and with more interactive conversational systems.

\newpage

\begin{acks}
This work was supported by the German Research Foundation (DFG) as part of the ``VACOS 2'' project (no. 388815326). We thank our study participants and our anonymous reviewers for their helpful feedback. We would also like to thank Jan Lattenkamp for his assistance with the technical implementation.
\end{acks}

\bibliographystyle{ACM-Reference-Format}
\bibliography{main}

\end{document}